%% file: Paper_20.tex
\documentclass[aps,prb,amssymb,showpacs,floatfix, twocolumn,superscriptaddress, bibliography]{revtex4-1}
\usepackage{amssymb,graphicx}
\usepackage[usenames,dvipsnames,svgnames,table]{xcolor}
\usepackage{epsfig,graphicx}
\usepackage{epstopdf}
\usepackage{amsmath}

\usepackage{lipsum}   
\usepackage{natbib}
\usepackage{braket} 

\usepackage[colorlinks=true]{hyperref}         
\hypersetup{
    bookmarks=true,         
    unicode=false,          
    pdftoolbar=true,        
    pdfmenubar=true,        
    pdffitwindow=false,     
    pdfstartview={FitH},    
    pdftitle={Probing trihedral corner entanglement for Dirac fermions},    
    pdfauthor={Author},     
    pdfsubject={Subject},   
    pdfcreator={Creator},   
    pdfproducer={Producer}, 
    pdfkeywords={keyword1} {key2} {key3}, 
    pdfnewwindow=true,      
    colorlinks=true,       
    linkcolor=magenta, 
    citecolor=JungleGreen,        
    filecolor=magenta,      
    urlcolor=black          
}

\newcommand{\Renyi}{R\'{e}nyi\ }

\newcommand{\eg}{{\it e.g.,}\ }
\newcommand{\ie}{{\it i.e.,}\ }
\newcommand{\X}{\gamma}

\newcommand{\beq}{\begin{equation}}
\newcommand{\eeq}{\end{equation}}
\newcommand{\beqa}{\begin{eqnarray}}
\newcommand{\eeqa}{\end{eqnarray}}
\newcommand{\beqar}{\begin{eqnarray*}}
\newcommand{\eeqar}{\end{eqnarray*}}
\newcommand{\reef}[1]{(\ref{#1})}

\newcommand{\cc}{\sigma}

\begin{document}
\title{Probing trihedral corner entanglement for Dirac fermions}

\author{Grigory Bednik}
\affiliation{Department of Physics and Astronomy, University of Waterloo, Ontario, N2L 3G1, Canada}

\author{Lauren E. Hayward Sierens}
\affiliation{Perimeter Institute for Theoretical Physics, Waterloo, Ontario N2L 2Y5, Canada}

\author{Minyong Guo}
\affiliation{Perimeter Institute for Theoretical Physics, Waterloo, Ontario N2L 2Y5, Canada}
\affiliation{Department of Physics, Beijing Normal University, Beijing 100875, P.R.~China}

\author{Robert C. Myers}
\affiliation{Perimeter Institute for Theoretical Physics, Waterloo, Ontario N2L 2Y5, Canada}

\author{Roger G. Melko}
\affiliation{Department of Physics and Astronomy, University of Waterloo, Ontario, N2L 3G1, Canada}
\affiliation{Perimeter Institute for Theoretical Physics, Waterloo, Ontario N2L 2Y5, Canada}

\date{\today}

\begin{center}
\begin{abstract}

We investigate the universal information contained in the \Renyi entanglement entropies for a free massless
Dirac fermion in three spatial dimensions.  Using numerical calculations on the lattice, we examine the case where 
the entangling boundary contains trihedral corners.  The entropy contribution arising from these corners grows logarithmically in the entangled subsystem's size with a universal coefficient. Our numerical results provide evidence that this logarithmic coefficient has a simple structure determined by two universal functions characterizing the underlying critical theory and the geometry of the corner. This form is similar to that of the analogous coefficient appearing for smooth entangling surfaces.
Furthermore, our results support the idea that one of the geometric factors in the corner coefficient is topological in nature and related to the Euler characteristic 
of the boundary, in direct analogy to the case of the smooth surface.
We discuss implications, including the possibility that one can use trihedral corner contributions to the \Renyi entropy to determine both of the universal central charges of the underlying critical theory. 

\end{abstract}
\end{center}

\maketitle

\section{Introduction}

Quantum critical systems possess universal quantities that describe the fixed points and renormalization group (RG) flows of the underlying field theory. Although the most familiar examples of such quantities are critical exponents extracted from the scaling of $n$-point functions,
universal numbers may also arise from the scaling behavior of entanglement entropies.
The latter crucially depends on the geometry of the bipartition boundary separating the two entangled regions of space.
For example, in $1+1$ space-time dimensions, the entanglement entropy of a critical system is known to scale logarithmically
in the length of the entangling region, with a coefficient proportional to the central charge $c$ of the underlying conformal field theory (CFT).
\cite{Holzhey_1994,Vidal_2002,Latorre_2003,Korepin_2004,Calabrese2004,Calabrese2009}
In higher dimensions, a rich variety of entangling geometries are possible, resulting in entanglement entropies that harbor 
many different universal features, yet to be fully explored.\cite{Solodukhin_2008,Casini_2009_1,Casini_2009_2,Kovacs_2012,Fursaev_2012,Witczak-Krempa_2016,Chojnacki_2016,Whitsitt_2016}
Of particular interest are universal quantities that decrease monotonically under RG flows;\cite{Myers:2010tj,Myers:2010xs, Casini:2018cxg,Casini:2012ei,Casini:2004bw,Casini:2017vbe,chm2,Lashkari:2017rcl}
these quantities are analogues to the central charge $c$, but for higher dimensions.  They can provide novel constraints to phase diagrams, as has recently been demonstrated for the case of certain strongly-interacting theories in 2+1 dimensions, of interest to condensed matter physics.\cite{Grover_2014}

The ground-state entanglement entropy (as well as the generalized \Renyi entropies) typically obeys an \textit{area law} such that,
up to leading order, it scales proportionally
to the surface area of the entangled subsystem.\cite{Sorkin_1983,Bombelli_1986,Srednicki_1993}
This contribution represents a power law divergence, which depends on the details of how the system is regulated at short-distances, \ie in the ultraviolet (UV).
Therefore the area law contribution does not contain universal information about the underlying critical theory.
However, subleading corrections to the area law can encode universal features of the theory 
that one can extract using various entanglement geometries.

\begin{figure}
\begin{center}
\def\svgwidth{0.63\columnwidth} 
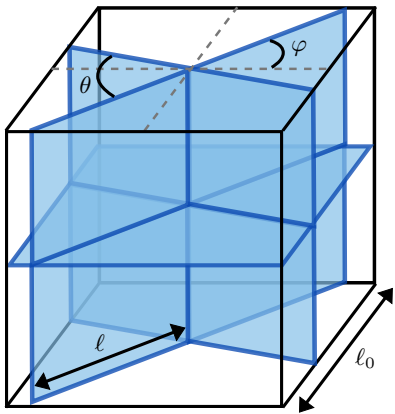
\end{center}
\caption{Divisions formed from three non-parallel planes that intersect in the center of a cube of length $\ell_0$.
These planes define various entanglement bipartitions $\{A,B\}$ with characteristic lengths $\ell$ that are used to extract trihedral corner coefficients.
In our calculations, we take the first plane to be parallel to the system's $xy$-plane, with the other two planes perpendicular to it.
The second plane is rotated by angle $\varphi$ relative to the system's $xz$-plane, 
and the third plane is rotated by angle $\theta$ relative to the second plane.
We consider $0 < \theta < \pi$. 
}
\label{fig:CubeCross-sections}
\end{figure}

One such universal quantity arises as the coefficient of a subleading logarithmic correction for entangling surfaces containing sharp vertices or corners. 
For a CFT in $2+1$ dimensions,
this universal logarithmic coefficient is a function of the corner angle 
and it can be used to obtain the central charge associated with the two-point function of the stress-energy tensor of the CFT.
\cite{Casini_2006,Casini_2009_1,Casini_2009_2,Hirata_2007,Kovacs_2012,Myers_2012,Kallin_2013,Kallin_2014,Helmes_2014,Devakul_2014_1,Stoudenmire_2014,Bueno_2015_1,Bueno_2015_2,Elvang_2015,Miao_2015,Bueno_2015_3,Sahoo_2016,Bueno_2016,DeNobili_2016,Helmes_2016}
For CFTs in $3+1$ dimensions, 
similar logarithmic corrections appear in the presence of trihedral corners (see Fig.~\ref{fig:CubeCross-sections}). 
However these logarithmic contributions are much less studied than their $(2+1)$-dimensional counterparts.

It is well-known in 3+1 dimensions that if the boundary of the entangled subsystem is curved, the \Renyi entropies will contain a
logarithmic correction.\cite{Solodukhin_2008, Fursaev_2012, Lee_2014} 
For a smooth entangling surface, the logarithmic coefficient is a linear combination of two  quantities sensitive to both the intrinsic and extrinsic geometry of the surface.\footnote{We are assuming that the background spacetime is flat.} Further, the prefactors in these geometric terms characterize the underlying critical theory and for the entanglement entropy, they reduce to the trace anomaly coefficients, $a$ and $c$ (see \eg Refs.~\onlinecite{Duff:1977ay,Duff:1993wm,birrell_davies}). Hence these two quantities characterizing the underlying CFT can be determined by evaluating entanglement entropies for subsystems with various geometries. 

In this paper we aim to explore to what extent the structure of the logarithmic coefficient discussed above for smooth entangling surfaces persists when the surface contains sharp folds and corners.
We study a system of free Dirac fermions in 3+1 dimensions, which is the simplest model for which the two prefactors can be isolated. Using finite-size lattice calculations, we numerically calculate the functional dependence of the trihedral corner coefficient on the \Renyi index and on a single opening angle.
Our results support the conclusion that the universal corner contribution is a linear combination of two independent terms, each of which contains two factors, one characterizing the critical theory and the other determined by the geometry. 
In particular, our results are consistent with the term proportional to the $a$ coefficient containing a topological factor related to the Euler characteristic of the entangling boundary.  Therefore we conclude that the trihedral corner, which is relatively simple to probe numerically (\eg much simpler than a spherical entangling surface), gives independent access to the two trace anomaly coefficients, $a$ and $c$, of the underlying critical theory.

\section{Entropy Scaling Theory for Dirac Fermions}
\label{sec:Theory}

The \Renyi entropy is a widely-used quantitative measure of quantum entanglement occurring in a physical system. 
For a system with a density matrix $\rho$ bipartitioned into two parts $A$ and $B$, the \Renyi entropy is defined as\cite{renyi1,renyi2}
\begin{equation}
S_{\alpha} = \frac{1}{1-\alpha}\, \log\mathrm{Tr}(\rho_A^{\alpha})\,,
\label{Salpha}
\end{equation}
where $\alpha$ is known as the \Renyi index and $\rho_A = \mathrm{Tr}_B \rho$ is the reduced density matrix for the subsystem $A$.  
In the limit where $\alpha \to 1$, this expression yields the von Neumann entanglement entropy.

When computing the \Renyi entropies in the vacuum state of a quantum field theory, the result is dominated by the short-distance correlations across the entangling surface and so the leading contributions have a geometric form. For example, evaluating $S_\alpha$ in a (3+1)-dimensional CFT for an entangling region $A$ with a smooth boundary $\Sigma$ yields,
\begin{equation}
S_\alpha(A)=B_{\alpha}\, \frac{{\mathcal A}_\Sigma}{\delta^2} +  
v_{\alpha}\,\log(\ell/\delta) + \cdots\, .
\label{EESmooth}
\end{equation}
The first term is the famous ``area law" contribution,\cite{Sorkin_1983,Bombelli_1986,Srednicki_1993} where ${\mathcal A}_\Sigma$ is the area of the entangling surface $\Sigma$ and $\delta$ is the short-distance cutoff. In the second term, $\ell$ is a characteristic length of the region $A$, and the ellipsis denotes contributions which are subleading to the logarithm. The first coefficient $B_\alpha$ is non-universal, \ie it depends on the details of the UV regulator. 
On the other hand,  $v_\alpha$ is regulator-independent and contains both geometric details about the entangling surface and universal data characterizing the underlying CFT such that\cite{Fursaev_2012,Lee_2014}$^,$\footnote{We assume here that the background geometry is flat, otherwise Eq.~\reef{EESmooth2} would contain a third term where the integrand is proportional to the Weyl curvature.}
\begin{equation}
 v_{\alpha}=-\int_\Sigma d^2\sigma\sqrt{h} \left[\frac{f_a(\alpha)}{2\pi}\,{\mathcal R}+\frac{f_b(\alpha)}{2\pi}\,{\widetilde K}^{{\hat \i}b}_a{\widetilde K}^{{\hat \i}a}_b
 \right]\, ,
\label{EESmooth2}
\end{equation}
where $\mathcal R$ in the intrinsic Ricci scalar and ${\widetilde K}^{\hat\i}_{ab}$ is the traceless part of extrinsic curvature. The two coefficients $f_{a,b}(\alpha)$ depend on the \Renyi index in a way that is characteristic of the underlying critical theory. In particular, they yield the trace anomaly coefficients with
$f_a(\alpha\to1)=a$ and $f_b(\alpha\to1)=c$. 
In this smooth-boundary case, the geometric factor corresponding to the $f_a(\alpha)$ term is topological since the integral of the Ricci scalar over the entangling surface yields twice its Euler characteristic. 
Further, this term can be isolated by choosing the entangling surface to be a sphere, in which case $v_{\alpha}=-4\,f_a(\alpha)$.\cite{Solodukhin_2008,Myers:2010tj,Myers:2010xs} 
We note that the corresponding trace anomaly coefficient $a$ is of particular interest since it plays a role analogous to the central charge in (1+1)-dimensional theories,\cite{Zamo} as it decreases monotonically under RG flows in 3+1 space-time dimensions.

In this paper, we consider \Renyi entanglement entropies $S_\alpha$ for entangling regions $A$ with characteristic length $\ell$ that are bounded by non-parallel planes in $3+1$ dimensions as in Fig.~\ref{fig:CubeCross-sections}. That is, rather than being smooth, the entangling surface contains sharp folds and corners, where two or three boundary planes intersect. 
General arguments\cite{Swingle_2010,HaywardSierens_thesis_2017} show that one expects $S_\alpha$ to scale as
\begin{equation}
S_{\alpha}(\ell) = B_{\alpha}\, \ell^2 + w_{\alpha}\, \ell + v_{\alpha}\,\log(\ell) + g_\alpha(\ell)\, ,
\label{EntropyForm}
\end{equation}
where for simplicity, we have set the UV cutoff (or lattice spacing) $\delta$ equal to one, 
that is, any factors of the length scale in the above expression can be interpreted as the dimensionless ratio $\ell/\delta$. In keeping with the previous discussion,  $B_{\alpha}$ and $w_{\alpha}$ are both regulator-dependent and $v_{\alpha}$ is a universal coefficient. 
The non-universal function $g_\alpha(\ell)$ is subleading to the logarithmic term and encodes all lower-order finite-size corrections that are proportional to non-positive powers of $\ell$ (or positive powers of the ratio $\ell_0/\ell$, where $\ell_0$ is the total system size).

From a geometric perspective, the leading term in Eq.~\eqref{EntropyForm} is the area law contribution.
The term proportional to $w_{\alpha}$ arises from the dihedral folds in the entangling surface, and it scales with the length of these folds.\cite{Klebanov_2012,Myers_2012}
The logarithmic term with coefficient $v_{\alpha}$ corresponds to the contributions to the \Renyi entropy coming from the trihedral corners.\cite{Casini_2009_1,TrihedralCornerScalar} 

As stated above, our goal in this paper is to examine to what extent the structure of the logarithmic coefficient $v_\alpha$  for a smooth entangling surface persists when the surface contains sharp folds and corners. In particular, our numerical calculations test the ansatz
\begin{equation}
v_{\alpha} = \beta_a\, f_a(\alpha) + \beta_b\, f_b(\alpha)\,,
\label{CoeffForm}
\end{equation}
where $f_{a,b}(\alpha)$ are the same universal functions of the \Renyi index as those appearing in Eq.~\reef{EESmooth2}. 
These universal functions characterize the underlying CFT.  For a given CFT,
the coefficients $\beta_{a,b}$ are assumed to depend only on the geometry (\ie the angles defining the trihedral corner) and to be independent of the \Renyi index. 

Further, we examine whether $\beta_a$ has a topological origin as in Eq.~\reef{EESmooth2}. 
That is, we test the idea that $\beta_a$ is given by $-2$ times the contribution of the trihedral corner to the Euler characteristic of the entangling surface. For example, the Euler character of a cube is $2$ (just as for a sphere) but the curvature is concentrated at the eight (identical) corners. 
Hence our proposal in this case is that one would should find $\beta_a = -{1}/{2}$ for each of the corners.

Previous numerical studies of the trihedral corner contribution have been made for the free\cite{TrihedralCornerScalar} and interacting\cite{Devakul_2014_2} scalar, restricted to the case of a cubic corner, \ie with $\theta=\pi/2$ in Fig.~\ref{fig:CubeCross-sections}. 
For a free (massless) scalar field, the universal functions obey $f_b({\alpha}) = {3} f_a({\alpha})$, and hence Ref.~\onlinecite{TrihedralCornerScalar} was only able to probe the particular combination $\left[ \beta_a+3\beta_b\right]_{\theta=\pi/2}$. 
Here, we turn to the case of Dirac fermions, where the $\alpha$-dependence of the two universal functions $f_{a,b}$ is instead given by\cite{Fursaev_2012,Lee_2014}
\begin{equation}
\begin{gathered}
f_a(\alpha) =  \frac{(\alpha+1)(37 \alpha^2+7)}{2880\alpha^3}\, ,\\
f_b(\alpha) = \frac{(\alpha+1)(17\alpha^2 + 7)}{960\alpha^3}\,.
\end{gathered}
\label{FunctionsAlpha}
\end{equation}
Setting $\alpha=1$ in these expressions again yields the trace anomaly coefficients of the Dirac fermion, \ie $f_a(\alpha=1) = 11/360 = a$ and $f_b(\alpha=1) = 1/20 = c$.\cite{birrell_davies,Fursaev_2012}$^,$\footnote{Note that with our present conventions, $a=\frac1{360}$ and $c=\frac1{120}$ for a free massless real scalar.} 
Combining Eqs.~\eqref{CoeffForm} and \eqref{FunctionsAlpha} yields the prediction
\begin{equation}
\frac{2880 \alpha^3 v_{\alpha}}{7(\alpha+1)}  = \frac{\alpha^2}{7} \left( 37 \beta_a + 51\beta_b \right) +\beta_a + 3 \beta_b\,,
\label{TransformedEqCoeffForm}
\end{equation}
which plays a central role in identifying the geometric coefficients $\beta_{a,b}$ in our numerical calculations. 

Further, we consider both cubic and non-cubic corners, as shown in Fig.~\ref{fig:CubeCross-sections}. That is, our general geometry is a trihedral corner where two of the planes are orthogonal to the first but they intersect each other with an opening angle $\theta$. Hence the cubic corner is recovered by setting $\theta=\pi/2$, but we also consider other configurations where $\theta\ne\pi/2$. 
As described above, in characterizing $\beta_a$ as topological, we are extending our application of (the $f_a(\alpha)$ portion of) Eq.~\reef{EESmooth2} to  entangling surfaces where the curvature is concentrated at trihedral corners.
Hence, for the general class of trihedral corners illustrated in Fig.~\ref{fig:CubeCross-sections}, we expect that \footnote{One can arrive this prediction, for example, by extending the previous arguments for a cube to a more general prism composed of identical corners. Such a prism has rectangular sides and end faces that are regular $n$-sided polygons. The corresponding opening angle for each trihedral corner is then $\theta=\pi(1-2/n)$. To generate a prediction for $\beta_a(\theta)$, we take $-2$ times the Euler character (which is 2) and divide by the total number of corners (which is 2$n$), thus arriving at $\beta_a(\theta)=-2/n$. Writing this expression in terms of $\theta$, we naturally arrive at Eq.~\reef{guess}.} 
\beq
\beta_a(\theta) = \frac{\theta}\pi-1,
\label{guess}
\eeq
such that $\beta_a$ depends linearly on the opening angle $\theta$. For the cubic corner (\ie $\theta=\pi/2$), this expression reduces to $\beta_a=-1/2$. 
We note that in the following, we do not directly probe $\beta_a(\theta)$ for single corners when $\theta\neq\pi/2$, but rather $\beta_a(\theta) + \beta_a(\pi-\theta)$ as explained in Sec.~\ref{sec:noncubic}.

We perform our numerical calculations for a system of free Dirac fermions 
on a three-dimensional cubic lattice with $N_x \times N_y \times N_z$ sites. 
We write the Hamiltonian in terms of the Hermitian gamma matrices 
and lattice momenta as
\begin{equation}
\begin{aligned}
H(k) &= \Gamma_1 \sin k_x  + \Gamma_2 \sin k_y + \Gamma_3 \sin k_z  \\
&\phantom{=\:} + \Gamma_0 B \left(3 - \cos k_x  - \cos k_y  - \cos k_z  \right),
\end{aligned}
\label{InitialHam}
\end{equation}
where the first three terms are conventional lattice Dirac terms and the last term is introduced to remove fermion doubling, which occurs due to extra
copies of the Dirac points arising at the time-reversal-invariant points at the boundary of the Brillouin zone 
(\textit{i.e.} the points with momentum components $0$ or $\pm \pi$).
We assume that the system is at half-filling such that all eigenstates with negative energies are filled, whereas all states with positive energies are empty. 
Such a system approaches a CFT in the low-energy limit when the number of lattice sites in each direction $N_x , N_y , N_z$ become infinitely large.

\section{Computational methods}
\label{sec:Methods}

To compute \Renyi entropy for various entangling geometries, we implement the scheme first presented in Ref.~\onlinecite{Peschel2003} (see also Refs.~\onlinecite{Casini_2009_2,Helmes} for more details). 
More specifically, we first 
find exact eigenvalues $w(k)$ and eigenvectors $v(k)$ of the Hamiltonian $H$. 
We then use these eigenvectors to compute a correlation matrix, which can be expressed as a sum of all outer products of the filled eigenstates such that
\begin{eqnarray}
C_{ij} = \langle c_i^\dagger c_j \rangle = \sum_{\text{filled states}} v_i(k) \, v_j^\dagger(k).
\label{Cmatrix}
\end{eqnarray}
Next, we restrict our consideration to the components $i$ and $j$ of $C$ that belong the subsystem $A$, thus obtaining a reduced correlation matrix $C_A$. 
Finally, following Ref.~\onlinecite{Casini_2009_2} (see also Ref.~\onlinecite{Helmes} for more details) we find the eigenvalues $\zeta$ of $C_A$ and compute the entanglement and \Renyi entropies according to
\begin{equation}
\begin{gathered}
S_1 = - \sum\limits_{\zeta} \left( \zeta \log\zeta + (1-\zeta)\log(1-\zeta) \right), \\ 
S_{\alpha} = \frac{1}{1-\alpha}  \sum\limits_{\zeta}\log \left( \zeta^{\alpha} + (1-\zeta)^{\alpha} \right).
\end{gathered}
\label{EntropiesEqs}
\end{equation}
After calculating \Renyi entropies for various geometries, we proceed to linearly combine them in order to isolate the logarithmic term in Eq.~\eqref{EntropyForm}. 
Specifically, we take the whole system to be a cube and intersect it by one, two or three planes as in Fig.~\ref{fig:CubeCross-sections}. 
As a result, we obtain a set of 26 subregions $A$ of three different kinds, 
\ie~6 subregions 
bordered by one crossing plane (type I), 
12 subregions  
bordered by two planes (type II), 
and 8 subregions  
bordered by three planes (type III). 
Only the figures of type III contain a trihedral corner.
Thus by computing the superposition of the entropies\cite{TrihedralCornerScalar}
\begin{equation}
\Delta S =
\sum_{i \in\,\text{type I}} S_i
- \sum_{i \in\,\text{type II}} S_i
+ \sum_{i \in\,\text{type III}} S_i,
\label{SumEntropies}
\end{equation}
it is possible to eliminate the quadratic and linear contributions to the entropy from the surfaces, dihedral angles and intersections with the system boundary. 
The surviving terms are the sum of all trihedral corner coefficients arising from the intersection of the three planes, up to subleading finite-size corrections. 
As discussed in Ref.~\onlinecite{TrihedralCornerScalar}, these 26 bipartitions contain redundant information, and therefore in practice we limit our consideration to 13 of them.
In the case when all planes are perpendicular to each other and parallel to the system's axes (see Fig.~\ref{fig:CubeCross-sections} with $\theta=\pi/2$ and $\varphi=0$), all trihedral corners are equivalent and thus the logarithmic coefficient is equal to the sum in Eq.~\eqref{SumEntropies} divided by the number of trihedral corners. 

We perform the above procedure with both open and periodic boundary conditions (OBCs and PBCs) imposed on the system. 
In both cases, we select the planes defining the bipartition to be parallel to the faces and passing through 
the center of the cubic system, which reduces the number of inequivalent bipartitions from 13 to 3. 
As a result of this choice for the locations of the planes, the ratio between the characteristic bipartition length $\ell$ and the total system size $\ell_0$ remains fixed -- see Fig.~\ref{fig:CubeCross-sections}.
In this way, subleading contributions within $g_\alpha(\ell)$ of Eq.~\eqref{EntropyForm} that scale as $\mathcal{O}(\ell/\ell_0)$ disappear upon performing the subtraction in Eq.~\eqref{SumEntropies}.
In the case of OBCs we diagonalize the lattice Hamiltonian numerically from the real-space representation obtained via a lattice Fourier transformation of the Eq. \eqref{InitialHam} . 
For PBCs we diagonalize the Hamiltonian analytically in momentum space and then numerically rewrite the eigenvectors $v_i(k)$ in coordinate representation before substituting them into Eq.~\eqref{Cmatrix}.
Both methods are constrained by the size of the reduced correlation matrix $C_A$, and numerically we find that the method of OBCs manifests the smallest subleading corrections.

In addition to the methods with three bipartitions for OBC and PBC,
we also compute the corner coefficient using a third method known as the
numerical linked-cluster expansion\cite{Rigol_2006,Rigol_2007_1,Rigol_2007_2,Tang_2013} (NLCE), 
as described in detail for corner entropy calculations in Refs.~\onlinecite{Kallin_2013,Kallin_2014,Kallin_thesis_2014, Helmes_2016, TrihedralCornerScalar}.
The NLCE can be applied to compute any extensive quantity $\mathcal{P}$, and in our case we take $\mathcal{P}$ to be the isolated trihedral corner contribution to the entropy summed over all possible locations of the corner in the lattice system.
In the NLCE, $\mathcal{P}$ is calculated for a lattice system by summing weights $W$ of finite systems (``clusters") $\cc$ such that
\begin{equation}
\mathcal{P} = \sum_\cc W(\cc),
\end{equation}
where the sum is in principle over all possible clusters $\cc$ (up to a given maximum size) that can be embedded in the total system.
As discussed in Refs.~\onlinecite{Kallin_2013,TrihedralCornerScalar}, one can estimate the quantity $\mathcal{P}$ by considering only cuboidal clusters $\cc$ such that the sum is over a much smaller number of terms.
The weights of each cluster are defined recursively such that
\begin{equation}
W(\cc) = \mathcal{P}(\cc) - \sum\limits_{s_\cc \in \cc} W(s_\cc),
\end{equation}
where $\mathcal{P}(\cc)$ is the desired quantity computed on the finite system $\cc$ and the sum is over all subclusters $s_\cc$ of the cluster $\cc$.

When using any of the three procedures described above (PBC, OBC and NLCE), 
we are interested in extracting the contribution to the \Renyi entropy from a single trihedral corner,
or, in the case where the planes are not mutually perpendicular, the contribution from an ``averaged" single corner as described in the following section.
Assuming that the total corner contribution to the entropy of a bipartition can be written as a sum of contributions from each corner
present on the boundary,
we divide the superposition of entropies in Eq.~\eqref{SumEntropies} by the total number of contributing corners. 
When we use 13 separate bipartitions to calculate Eq.~\eqref{SumEntropies}, for the OBC and NLCE methods such superpositions contain contributions from four trihedral corners,
whereas for the PBC calculation the superposition of entropies contains contributions from 32 trihedral corners.

\section{Results}
\label{sec:Results}

Using the techniques described in the last section, we extract the corner coefficient $v_\alpha$ for various trihedral angles from numerical calculations of the superposition of entropies in Eq.~\eqref{SumEntropies}. 
We use data from several finite-size systems to extrapolate $v_\alpha$ to the infinite size limit, enabling us to make comparisons to results from continuum theory. 
From the trihedral corner coefficients, we extract the coefficients $\beta_a$ and $\beta_b$ corresponding to both cubic and non-cubic trihedral angles. 

\subsection{Finite-size methodologies}
\label{FSS}

\begin{figure}[t]
\includegraphics{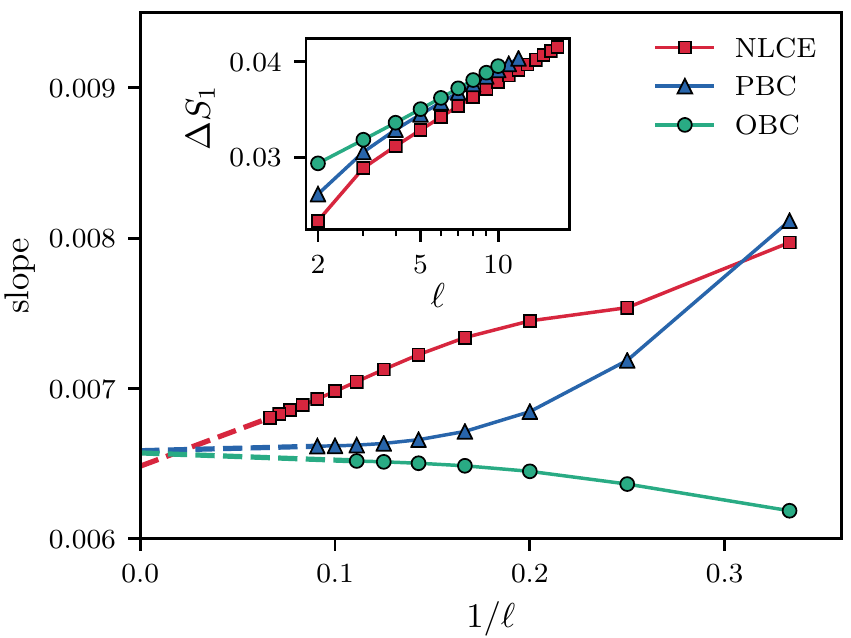}
\caption{The slope of the superposition of von Neumann entropies $\Delta S_1$ as a function of the inverse of the characteristic length scale $\ell$.
The inset has the same legend as the main plot, and shows $\Delta S$ as a function of $\ell$ using three different methods to compute the entropy as discussed in the main text.
In the main plot, we show the pairwise slope of the data from the inset. 
The dashed lines illustrate extrapolations to the thermodynamic limit used to obtain $v_1$.
}
\label{fig:slope}
\end{figure}

To perform the extrapolation of the corner coefficient to the infinite size limit, we begin by
considering the dependence of the superposition of entropies on the system size.
For the OBC and PBC calculations, we take the characteristic size to be that of the subsystem $\ell$ while, for the NLCE, this size is the so-called ``order'' of the NLCE, which corresponds to the maximum linear size of the considered cuboid clusters.\cite{Sahoo_2016,TrihedralCornerScalar}
In Fig.~\ref{fig:slope} (inset) we present an example of this size dependence for $\alpha =1$. 
The deviation from linearity in this graph indicates that the superposition contains additional corrections past the logarithmic form expected from the trihedral corner.
These corrections are caused by a variety of finite-size effects, such as the finite ratio of system and subregion sizes, and 
(for $\alpha>1$) the conical singularity of the multi-sheeted Riemann surface.\cite{Cardy_2010,Sahoo_2016}
To attempt to account for these corrections, we perform a two-step fit.\cite{Sahoo_2016,TrihedralCornerScalar}
In the first step, we compute the slope for each neighboring pair of points on the graph of $\Delta S$ versus $\log \ell$, which is a finite-size estimate for the corner coefficient. 
In the second step, we plot each estimate as a function of the inverse system size (see Fig. \ref{fig:slope}), and extrapolate to the thermodynamic limit ($1/\ell \to 0$). 
The extrapolation consists of a linear fit to the three points corresponding to the largest length scales.

Having extracted values for the trihedral corner coefficients $v_{\alpha}$ in this manner, we can now proceed to analyze them 
as a function of the \Renyi index $\alpha$ for several angles 
$\theta$ between the planes. 
We begin in the next section with the case of the cubic trihedral corner.

\subsection{Cubic trihedral angles}
\label{sec:cubic}

\begin{figure}[tb]
\centering
\includegraphics{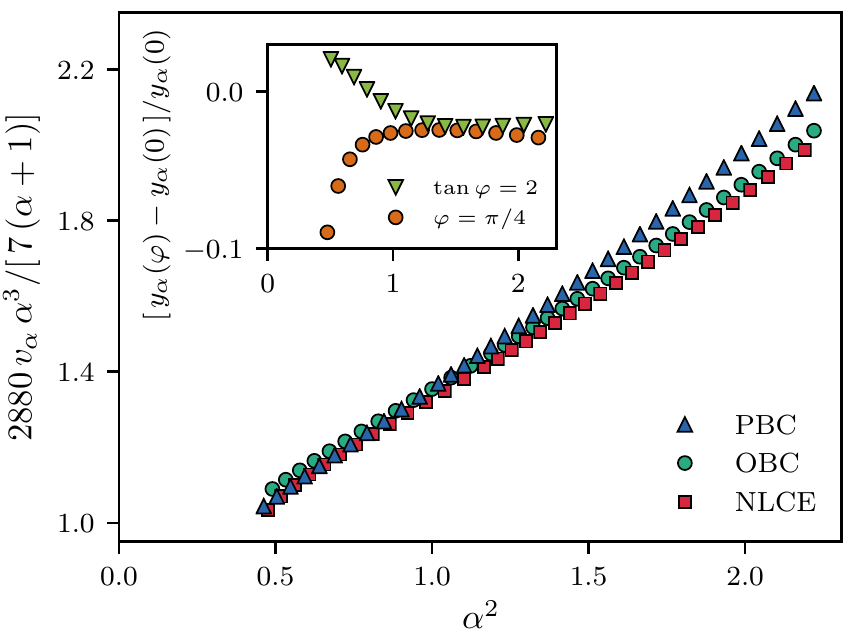}
\caption{The $\alpha$-dependence of the cubic ($\theta=\pi/2$) trihedral corner coefficient for Dirac fermions,
as extracted using three different methods (with $\varphi=0$).
The inset illustrates how the OBC results vary when the mutually-perpendicular planes are rotated by angle $\varphi$ about the system's $z$-axis 
(see Fig.~\ref{fig:CubeCross-sections}).
The inset has the same horizontal axis as the main plot and uses the definition $y_\alpha = 2880\, v_\alpha \alpha^3 / [7(\alpha+1)]$. 
}
\label{fig:Result_linear}
\end{figure}

We start by considering the case of cubic trihedral angles with planes parallel to the system's axes such that $\theta = \pi/2$ and $\varphi=0$ in Fig.~\ref{fig:CubeCross-sections}.
Using our numerical extrapolation for $v_{\alpha}$ described in the last section, we compute the left-hand-side of 
Eq.~\eqref{TransformedEqCoeffForm} and plot it as a function of $\alpha^2$, as illustrated in Fig.~\ref{fig:Result_linear}. 
One can see that the numerical graphs are approximately linear, as predicted by Eq.~\eqref{TransformedEqCoeffForm}. 

Next, we extract the parameters of the best-fit line $L\, \alpha^2 + C$ to the data in Fig.~\ref{fig:Result_linear}, obtaining
\begin{equation}
	\begin{gathered}
	L = 0.58 \pm 0.07, \\
	C = 0.77 \pm 0.08.
\label{CoeffNumbers}
	\end{gathered}
\end{equation}
Here, the uncertainties of the fit parameters represent the discrepancy between the different finite-size methods (described in Section \ref{sec:Methods}),
averaged for the different $\alpha$ within the range $(0.67, 1.5)$. 
Using Eq.~\eqref{TransformedEqCoeffForm}, these parameters allow us to find the coefficients $\beta_{a, b}$, yielding 
\begin{equation}
	\begin{gathered}
	\beta_a = -0.45 \pm 0.10, \\
	\beta_b = 0.41 \pm 0.06.
	\end{gathered}
\end{equation}
From here, we see that the numerically-extracted value of $\beta_a$ is consistent with $-{1}/{2}$, 
which is the prediction from the conjecture that this geometric coefficient is given by $-2$ times the contribution of the corner to the Euler character of the entangling surface.

In the next section, we explore other (non-cubic) trihedral angles with $\theta \ne\pi/2$.  
Such angles require one or more of the planes in Fig.~\ref{fig:CubeCross-sections} to no longer be perpendicular to the underlying cubic lattice that is the UV regularization of the theory.  
Before we explore different $\theta$, we first confirm that a misalignment of the entangling boundary with the lattice regularization does not produce any pathological effects
or scaling ambiguities.\cite{chm2}  
To do so we rotate two of the planes intersecting the cube in such a way that the trihedral angles are kept cubic 
(see Fig.~\ref{fig:CubeCross-sections} with $\theta=\pi/2$ and $\varphi \neq 0$). 
We present the results for the difference between the left-hand side $y_\alpha$ of Eq.~\eqref{TransformedEqCoeffForm} for $\varphi$ and $\varphi=0$ in the inset of Fig.~\ref{fig:Result_linear}.
We see that $y_\alpha$ has the largest discrepancies for $\alpha < 1$, 
but that it
differs by at most $10\%$ 
 from 
its $\varphi=0$ value for the range of $\alpha$ studied.
Repeating our analysis, all of the resulting coefficients of the fit fall into the range of Eq.~\ref{CoeffNumbers} within errors. 
We thus have confidence that our results are not significantly affected by a misalignment of the entangling boundary with the lattice regularization, and  
we proceed to explore other angles $\theta$ below.

\subsection{Non-cubic trihedral angles}
\label{sec:noncubic}

In the previous section we presented evidence that the trihedral corner coefficient can be used to numerically extract the geometric coefficients  $\beta_{a,b}$. In particular, we found for the $\theta=\pi/2$ corner that $\beta_a = -{1}/{2}$ to a high degree of accuracy, which is consistent with our conjecture that $\beta_a$ is topological. 
We now further test this hypothesis by exploring cases where the trihedral angles are non-cubic and the conjectured form of $\beta_a$ is given by Eq.~\reef{guess}. 
Specifically, we study geometries in Fig.~\ref{fig:CubeCross-sections} with $\theta \neq \pi/2$ (and $\varphi=0$), summing the contributions of all trihedral corners and dividing by the total number of corners that survive in Eq.~\eqref{SumEntropies}. 
In these non-cubic cases, the sum over the type III subregions (which are bordered by three planes) includes contributions from two non-equivalent types of corners with angles $\theta$ and $\pi - \theta$, and we thus introduce the average logarithmic coefficient,
\begin{equation}
\overline{v}_{\alpha} = \frac{1}{2} \left[ v_\alpha(\theta) + v_\alpha(\pi - \theta) \right]\,.
\label{eq:v_bar}
\end{equation}
We note that $\overline{v}_{\alpha} = v_{\alpha}$ for the cubic ($\theta = \pi/2$) case discussed in Sec.~\ref{sec:cubic}.
We also introduce the notation that this $\overline{v}_{\alpha}$ is related through Eq.~\eqref{TransformedEqCoeffForm} to the averaged geometric coefficients 
\begin{equation}
\overline{\beta}_{a,b} = \frac{1}{2} \left[ {\beta}_{a,b}(\theta) + {\beta}_{a,b}(\pi - \theta) \right]\, .
\label{eq:beta_bar}
\end{equation}
If the geometric factor $\beta_a$ has a topological character as proposed in Eq.~\reef{guess}, then in fact we should find that $\overline{\beta}_a=-1/2$ irrespective of the value of $\theta$.
We emphasize that, with the procedure described in Sec.~\ref{sec:Methods}, we are able to extract the averages $\overline{v}_{\alpha}$ and $\overline{\beta}_{a, b}$, 
but we do not have a means of accessing individual $v_\alpha(\theta)$ and $\beta_{a, b}(\theta)$ for general $\theta \neq \pi/2$.

\begin{figure}
\centering
\includegraphics{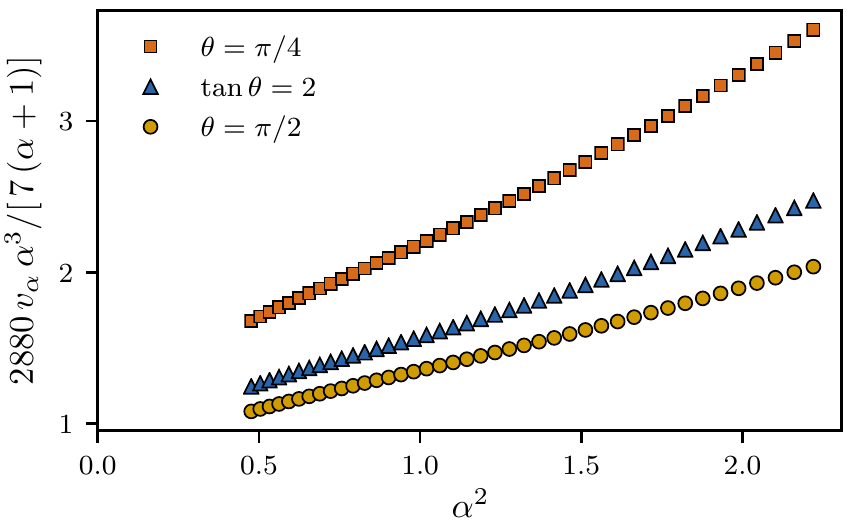}
\caption{The $\alpha$-dependence of the trihedral corner coefficient for Dirac fermions for different angles $\theta$ (see Fig.~\ref{fig:CubeCross-sections}). 
Here all calculations have $\varphi=0$ and are performed on systems with OBCs.
}
\label{fig:Result_non-cubic}
\end{figure}

We perform calculations of the logarithmic corner coefficient for angles $\theta=\pi/4$ and $\theta = \arctan(2)$.  
For these calculations we use a system with OBCs since we have already confirmed that this method gives results consistent with other finite-size methods for $\theta=\pi/2$ and $\varphi \neq 0$.
We perform the same finite-size extrapolation and fitting as in the previous section, and obtain the 
results shown in Fig.~\ref{fig:Result_non-cubic}.  
From this plot, 
we extract the values for $\overline{\beta}_a$ and $\overline{\beta}_b$ given in Table~\ref{tab:fitResults} for the different trihedral angles.
We see in each case that the range of values for $\overline{\beta}_a$ is consistent with $-{1}/{2}$, in agreement with our 
conjecture about the topological nature of this coefficient. In contrast, we find that $\overline{\beta}_b$ varies in a nontrivial way with $\theta$.

In the next section, we would like to compare our results with other analyses for the Extensive Mutual Information (EMI) model~\cite{eminotes,wilfor} and for the free scalar.\cite{TrihedralCornerScalar} 
However, in these models, it is not possible to extract $\overline{\beta}_{a, b}$ simultaneously, as we do for the free fermion. 
Instead, in these models, $\overline{\beta}_b$ can be extracted only after assuming prior knowledge of $\overline{\beta}_a$. 
Therefore, to compare our results on equal footing, we redo the fits in the last two columns of Table~\ref{tab:fitResults}, with the conjectured topological coefficient fixed to be $\overline{\beta}'_a = -1/2$. Our results again confirm that the corresponding coefficient $\overline{\beta}'_b$ varies in a nontrivial way with $\theta$, which we examine and compare against the other aforementioned analyses below.

\begin{table}
\setlength\extrarowheight{2pt}
\begin{center}
\begin{tabular}{|c||c|c|c|c||c|c|}
\hline
$\theta$ 	& $L$ 	& $C$  & $\phantom{\Big(} \overline{\beta}_a \phantom{\Big)}$ 	& $\overline{\beta}_b$ & $\overline{\beta}'_a$ & $\overline{\beta}'_b$   \\
\hline
$\pi/2$ 		 	 & 0.54(4) & 0.80(5) & --0.49(5) & 0.43(3) & --$\frac{1}{2}$ & 0.438(1)  \\[2pt]
\hline
$\mathrm{tan}\theta\!=\!2$ 
 & 0.68(12) & 0.87(15) & --0.50(17)  & 0.46(11) & --$\frac{1}{2}$ & 0.462(5)  \\[2pt]
\hline
$\pi/4$		 	 & 1.08(12) & 1.07(15) & --0.53(17) & 0.54(11) & --$\frac{1}{2}$ & 0.524(8) \\[2pt]
\hline
\end{tabular}
\end{center}
\caption{ 
Columns 2 and 3 correspond to fitting a straight line $L \alpha^2 + C$ to the data in Fig.~\ref{fig:Result_non-cubic}, 
where $L$ and $C$ respectively represent the slope and intercept of this fitted line. 
Columns 4  and 5 show the corresponding
$\overline{\beta}_a$ and $\overline{\beta}_b$,
which are the averaged coefficients as predicted from Eq.~\eqref{TransformedEqCoeffForm}. 
Finally, columns 6 and 7 show the result for these coefficients, when the fit is redone fixing $\overline{\beta}'_a=-1/2$ 
as would be required if it has the topological character described in the main text. The results (and uncertainties) are determined using OBCs only.
}
\label{tab:fitResults}
\end{table}

\subsection{Analysis of $\beta_b$}

\begin{figure}
\centering
\includegraphics{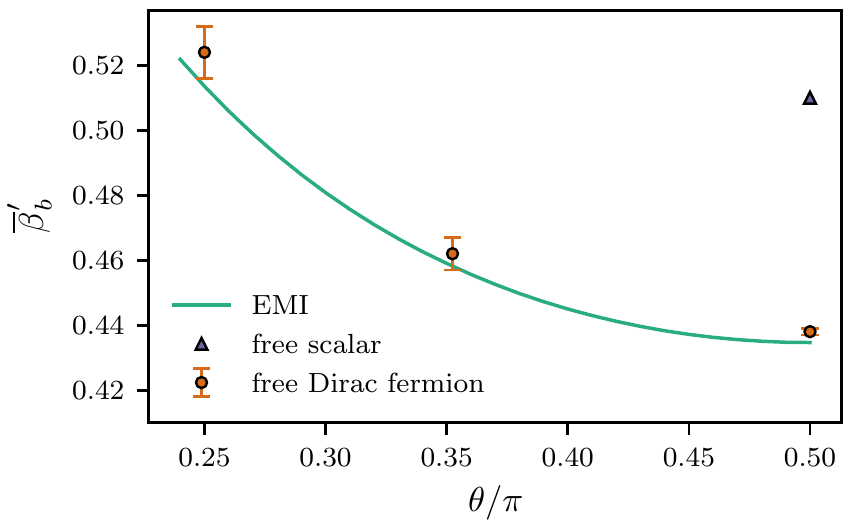}
\caption{The (averaged) geometric factor $\overline{\beta}'_b$ as a function of the opening angle $\theta$, 
determined assuming that $\overline{\beta}'_a=-1/2$ as stipulated by Eq.~\reef{guess}. We compare our results for the Dirac fermion with results for the EMI model~\cite{eminotes} and for the free scalar at $\theta=\pi/2$.\cite{TrihedralCornerScalar} 
}
\label{fig:compare-betab}
\end{figure}
%
Let us now comment on our results for the other geometric factor $\beta_b$.  Our first observation from Table~\ref{tab:fitResults} is that when this coefficient is averaged as in Eq.~\reef{eq:beta_bar}, the result has the opposite sign compared $\overline{\beta}_a$. Further,  $\overline{\beta}'_b$ clearly exhibits a nontrivial dependence on $\theta$, again in contrast to $\overline{\beta}'_a$. Hence we do not expect that $\beta_b$ has a topological character like $\beta_a$. In Fig.~\ref{fig:compare-betab}, we compare our results for a Dirac fermion to the analogous results for two other theories. The first comparison is with a massless free scalar, for which Ref.~\onlinecite{TrihedralCornerScalar} provides the single data point: $\overline{\beta}'_b(\theta=\pi/2) \simeq +0.51$.  
As with the results in Table \ref{tab:fitResults}, the prime here indicates that this result for the scalar was  determined by assuming that $\overline{\beta}'_a=-1/2$, in keeping with the hypothesis that $\beta_a$ is described by Eq.~\reef{guess}. 
The second comparison is with the EMI
model.\cite{Casini:2005rm,Casini:2008wt,Swingle_2010} As its name suggests, this model is characterized by the special property that the
mutual information $I(A,B)$ is extensive, \ie $I(A,B \cup C) = I(A,B) + I(A,C)$.\footnote{Recall the mutual information of two (non-intersecting) regions $A$ and $B$ can be defined in terms of the entanglement entropies as $I(A,B)=S(A) + S(B) - S(A\cup B)$.} 
We stress that the EMI model does not correspond to an explicit conformal field theory. 
Rather, this model provides a simple framework in which to evaluate the entanglement entropy of various geometric configurations, 
which can then be used as an interesting benchmark to compare with the results for specific CFTs and specific configurations 
(see, for example, Refs.~\onlinecite{Bueno_2015_1,Bueno_2015_2,Witczak-Krempa_2016}). 
The results for $\overline{\beta}'_b(\theta)$ for the EMI model are shown in the figure, with the details of the corresponding calculations to appear elsewhere.\cite{eminotes}

Examining Fig.~\ref{fig:compare-betab}, we see that for both the EMI model and the Dirac fermion, $\overline{\beta}'_b$ decreases as $\theta$ increases over the range shown and reaches a minimum at $\theta=\pi/2$. In fact, the EMI calculations indicate that this quantity diverges as $\theta$ approaches zero. 
At $\theta=\pi/2$, the EMI and fermion values nearly coincide, \ie the fermion value given in Table~\ref{tab:fitResults} (for the OBC finite-size method) is less than 1\% away from the EMI result.
Furthermore, if we instead roughly estimate the error in the fermion result by additionally accounting for the discrepancy between the three different finite-size methods, 
then we find that the fermion and EMI results are actually consistent within this aggregate error. 
As shown in the figure, $\overline{\beta}'_b(\theta)$ agrees quite well for these two theories across the full range of angles shown there.
In contrast,
the scalar value is about 17\% higher at $\theta=\pi/2$ and is clearly distinct from the other two theories.
We might speculate that the reason that the fermions and the EMI model are in close agreement, but not the scalar, is related to ratio of the trace anomaly coefficients in these theories, \ie $c/a=3$, $18/11$ and $3/2$ for the scalar, fermion and EMI model,\cite{eminotes,wilfor} respectively. That is, this ratio is almost the same for the two theories where 
$\overline{\beta}'_b(\theta)$ shows good agreement 
but larger by (roughly) a factor of two in the scalar field theory. 
We expect that for different CFTs, $\beta_b$ will show the same basic trends of decreasing with increasing $\theta$, but that the detailed values and structure will be distinct. 

The behavior of $\overline{\beta}'_b(\theta)$ is reminiscent of that found for the universal corner coefficient appearing in the entanglement entropy of CFTs in 2+1 dimensions.\cite{Bueno_2015_1} In order to compare different theories, the corner coefficient was normalized by the central charge $C_T$, which defines the two-point function of the stress tensor in the particular CFT of interest. With this choice, the corresponding geometric factor showed the same overall behavior in different theories, while differing in the detailed structure. 
However, it is striking how similar the geometric factor remains for all theories. For example, the functions for a free scalar (\ie one degree of freedom and zero coupling) and for a holographic CFT (\ie a large number of degrees of freedom and strong coupling) differed by no more than 13\% across the full range $0\leq\theta\leq\pi$.

It is prudent to recall that in extending the above discussion to general \Renyi entropies, the universal corner coefficient 
for $(2+1)$-dimensional CFTs did not factorize in a simple way.\cite{Bueno_2015_3} 
That is, the dependence of this corner function on the opening angle is (slightly) different for different values of the \Renyi index. 
Further, this non-factorization was already apparent in examining the corner coefficient for free massless scalars and fermions 
in 2+1 dimensions. 
One might see
these observations
as being in tension with our ansatz in Eq.~\reef{CoeffForm} and numerical findings for the trihedral corner coefficient for $(3+1)$-dimensional CFTs. 
However, with two independent terms, Eq.~\reef{CoeffForm} is more sophisticated than the simple ansatz $a_\alpha(\theta) = f(\alpha)\,a_1(\theta)$ ruled out for $2+1$ dimensions in Ref.~\onlinecite{Bueno_2015_3}. That is, $v_\alpha$ does in fact have a different $\theta$ dependence for different values of $\alpha$. Hence there is no real disagreement with the findings for 
$2+1$ dimensions.

\section{Discussion}
\label{sec:Conclusions}

In this paper, we have studied the \Renyi entanglement entropies of Dirac fermions in three spatial dimensions, using finite-size lattice calculations. In our numerical studies, we considered regions that were bounded by non-parallel planes and we focused on the universal coefficient $v_\alpha$ of the logarithmic contribution in Eq.~\reef{EntropyForm}, which arises due to a trihedral corner in the entangling boundary.
We aimed to test two conjectures with regard to this trihedral corner coefficient. 
First, we examined whether the functional dependence on the \Renyi index $\alpha$ and on the geometry is described by Eq.~\reef{CoeffForm} such that
$v_{\alpha} = \beta_a\, f_a(\alpha) + \beta_b\, f_b(\alpha)$.
Here $f_{a,b}(\alpha)$ are the same universal coefficients as those controlling the log coefficient in Eq.~\reef{EESmooth2} for smooth entangling surfaces. 
The coefficients $\beta_{a,b}$ are assumed to depend only on the geometry (\ie the angles defining the trihedral corner) and to be independent of the \Renyi index. 
Our second test was to examine 
whether the $\beta_a$ has the same topological character as it does in Eq.~\reef{EESmooth2} for smooth boundaries. 
That is, we tested the conjecture that $\beta_a$ is given by $-2$ times the contribution of the trihedral corner to the Euler characteristic of the entangling surface. 

Substituting the fermion coefficients in Eq.~\reef{FunctionsAlpha} into the ansatz of Eq.~\reef{CoeffForm} yields a simple $\alpha$ dependence for the quantity $y_\alpha = 2880\, v_\alpha \alpha^3 / [7(\alpha+1)]$, as shown in Eq.~\reef{TransformedEqCoeffForm}. 
Fig.~\ref{fig:Result_non-cubic} (as well as Table \ref{tab:fitResults}) shows that $y_\alpha=L\alpha^2 +C$ provides a good fit for our numerical results for the corner coefficient for all three choices of $\theta$ in the vicinity of $\alpha=1$. 
This result provides support for the functional form of the ansatz in Eq.~\reef{CoeffForm}. 
Further, using Eq.~\reef{TransformedEqCoeffForm} to translate these fits to expressions for the averaged geometric coefficients in Eq.~\eqref{eq:beta_bar},
we find that the best fit values of $\overline{\beta}_{a}$ are very close to $-1/2$ in all three cases. 
Hence these fits support the conjecture that $\beta_a$ is topological and given by Eq.~\eqref{guess}.
Therefore, we find strong support for both of our conjectures for the structure of the log coefficient $v_\alpha$.

However, it is clear that our ansatz in Eq.~\reef{CoeffForm} should be subjected to  further tests before being fully accepted. 
In fact, a number of independent tests to extend the present work are possible. 
For example, we can investigate cases where the trihedral corner has an opening angle greater than $\pi$. 
To do so, recall that in Fig.~\ref{fig:CubeCross-sections}, we have divided the cube into eight subregions, 
each of which possesses a corner with angle $\theta$ or $\pi-\theta$, for $0<\theta<\pi$. The terms corresponding to these eight subregions then appear in the type III sum in Eq.~\reef{SumEntropies}. By combining these subregions in triplets above and below the $xy$-plane,
we can instead form  eight type III  regions corresponding either to an opening angle $2\pi-\theta$ or $\pi+\theta$, for $0<\theta<\pi$.  
In order to isolate the logarithm contribution, in this case we also change the sign of some of the type I terms; specifically those for which the entangling geometry borders the $xy$-plane, due to the fact that the type III regions cover the $xy$-plane three times in this case.
The coefficient of the resulting log contribution then corresponds to 
\begin{equation}
\widetilde{\overline{v}}_{\alpha} = \frac{1}{2} \left[ v_\alpha(2\pi-\theta) + v_\alpha(\pi + \theta) \right]\,,
\label{eq:v_bar2}
\end{equation}
with $0 < \theta < \pi$. 
The contribution of these corners to the Euler character is positive and thus the sign of the $f_a(\alpha)$ term in Eq.~\eqref{CoeffForm} should be reversed compared to the present calculations. 
Specifically, one expects to find
$\widetilde{\overline{\beta}}_{a} = \beta_a(\theta = 3\pi/2) = +1/2$ from Eq.~\reef{guess}.
Other interesting choices for the type III regions (along with appropriate adjustment of the coefficients of the terms in Eq.~\reef{SumEntropies}) yield configurations where the term proportional to $f_a(\alpha)$ vanishes according to our ansatz. 

Another simple test would come from extending the calculations of  Ref.~\onlinecite{TrihedralCornerScalar} for a massless free scalar to new values of the opening angle $\theta$. 
Recall that the numerical calculations there found that $v_\alpha$ was roughly proportional to the expected scalar function $f_b({\alpha}) = {3} f_a({\alpha})$, but only considered $\theta=\pi/2$. 

Finally, let us explore the implications of our conjectures that the corner coefficient $v_\alpha$ takes the form given in Eq.~\reef{CoeffForm} with $\beta_a$ given by Eq.~\reef{guess}. 
Recall that the functions $f_a$ and $f_b$ are known to reduce to the trace anomaly coefficients $a$ and $c$ of the underlying CFT when the \Renyi index is $\alpha=1$. 
Our results offer the tantalizing possibility that one might be able to gain information about these quantities 
simply by examining entanglement geometries formed from intersecting planes with local trihedral corners. 
To explore this, we evaluate Eq.~\reef{CoeffForm} for two different angles at $\alpha=1$ and average as in Eq.~\reef{eq:beta_bar},  to produce the ratio
\beq
\X\equiv\frac{\overline{\beta}_b(\theta_1)}{\overline{\beta}_b(\theta_2)}=
\frac{\overline{v}_1(\theta_1)+a/2}{\overline{v}_1(\theta_2)+a/2}\,.
\label{one}
\eeq
Since the geometric factor $\beta_b$ is independent of the \Renyi index, the ratio in Eq.~\reef{one} can also be determined by considering the $\alpha$ derivative of  Eq.~\reef{CoeffForm}.
In particular, we find
\beq
\X=
\frac{\partial\overline{v}_1(\theta_1)+\frac12 \partial_\alpha f_a (\alpha)\big|_{\alpha=1}}{\partial\overline{v}_1(\theta_2)+\frac12\partial_\alpha f_a (\alpha)\big|_{\alpha=1}}\,,
\label{twoX}
\eeq
where our notation is $\partial\overline{v}_1=\partial_\alpha \overline{v}_\alpha|_{\alpha=1}$. However, we can simplify the latter expression with the relation\cite{Perlmutter:2013gua}
\beq
\partial_\alpha f_a (\alpha)\big|_{\alpha=1}=-\frac12\, f_b(\alpha=1)=-\frac{c}2\,,
\label{useful}
\eeq  
which, upon substitution into Eq.~\reef{twoX}, yields
\beq
\X=
\frac{\partial\overline{v}_1(\theta_1)-c/4}{\partial\overline{v}_1(\theta_2)-c/4}\,.
\label{two}
\eeq
Combining Eqs.~\eqref{one} and~\eqref{two} now generates the equation
\begin{eqnarray}
\frac{a}{2} \left( \partial \bar v_1(\theta_1) - \partial \bar v_1 (\theta_2)\right)
+ \frac{c}{4} \left(\bar v_1(\theta_1) -\bar v_1(\theta_2) \right)
\nonumber\\
= \bar v_1(\theta_1) \partial \bar  v_1 (\theta_2) - \bar  v_1(\theta_2) \partial \bar  v_1 (\theta_1)\,,
\label{key}
\end{eqnarray}
involving the two central charges. 

A first approach to utilizing Eq.~\eqref{key} is to treat both $a$ and $c$ as unknown variables.
One could numerically evaluate $\bar v_{1}$ and $\partial \bar v_{1}$ for at least three angles and 
then solve for $a$ and $c$ by determining the intersection of any two of the resulting equations. 
Repeating the calculations described in this paper for many angles would allow us to extract the charges $a$ and $c$ by fitting the resulting numerical data to Eq.~\eqref{key} for many different pairs. 

There is an interesting consistency check of this numerical approach.
Consider the case of three angles $\theta_{1,2,3}$ and hence three distinct copies of Eq.~\eqref{key}. For each of these equations, the left-hand side is linear in $\bar v_1$ and $\partial \bar  v_1$, while the right-hand side is quadratic.
Upon adding the three equations together (with appropriate signs), the left-hand side vanishes but the right-hand side does not, and we are left with a nontrivial (quadratic) equation for these quantities. Hence we have a consistency test for these numerically determined quantities, which can be expressed as 
\begin{eqnarray}
0&=&\frac{\partial \bar  v_1 (\theta_2)}{\partial \bar  v_1 (\theta_1)}-\frac{\partial \bar  v_1 (\theta_3)}{\partial \bar  v_1 (\theta_1)}
-\frac{\bar  v_1(\theta_2)}{\bar  v_1(\theta_1)}
+\frac{\bar  v_1(\theta_3)}{\bar  v_1(\theta_1)}
\nonumber\\
&&\quad+\frac{\bar  v_1(\theta_2)\partial\bar  v_1 (\theta_3) }{\bar  v_1 (\theta_1)\partial \bar  v_1 (\theta_1)} -\frac{\bar v_1(\theta_3) \partial \bar  v_1 (\theta_2)}{\bar  v_1 (\theta_1) \partial \bar  v_1 (\theta_1)}\,. 
\label{test}
\end{eqnarray}
Here, we have normalized the expression by dividing by $\bar  v_1 (\theta_1)\,\partial \bar  v_1 (\theta_1)$ so that each term above is $O(1)$. 
We expect that combining the numerical results on the right-hand side will yield a small answer (rather than precisely zero), which would be a nontrivial confirmation of the hypotheses being examined in this paper.

A second approach to working with Eq.~\reef{key} arises if we already have access to one of the central charges by other means, \eg determining $c$ from the two-point function of the stress tensor. In this case, we can determine the other charge from one copy of the equation, \ie after obtaining $\bar v_{1}$ and $\partial \bar v_{1}$ numerically for two angles. For example, given $c$, we can evaluate the central charge $a$ as
\beq
a=2\,
\frac{\overline{v}_1(\theta_1)-\X\,\overline{v}_1(\theta_2)}{\X-1}\,,
\label{one2}
\eeq
where $\X$ would be evaluated using Eq.~\reef{two}.
Again, evaluating $\bar v_{1}$ and $\partial \bar v_{1}$ for more than two angles would  presumably yield a more accurate evaluation of $a$. 

Before closing, we comment on applying the above approaches to our numerical results for the free fermion. In particular, it is straightforward to evaluate $\bar v_1$ and $\partial \bar  v_1$ for the three opening angles  $\theta=\pi/2, \arctan(2)$ and $\pi/4$. 
Turning first to the consistency test presented in Eq.~\reef{test}, we find that the right-hand side is $0.002(37)$ for our OBC results, in good agreement with the expected vanishing. 
The different versions of Eq.~\reef{key} constructed with the three distinct pairings of the angles define different lines in the ($a$,$c$)-plane, but we find for our data that these lines are essentially degenerate within the error bars produced by the numerical uncertainties. 
Hence this approach will only produce reliable estimates of $a$ and $c$ with more accurate evaluations of $v_\alpha$.
This sensitivity may also be reduced somewhat if we consider a wider range of opening angles, \eg angles close to $3\pi/2$ as suggested above.

We can also apply the second approach to our data, \eg by assuming the known value $c=1/20$. 
In this case, we find that Eq.~\reef{one2} yields a best-fit value for the other central charge of $a = 0.027(4)$, which is consistent with the known value $a=11/360 \approx 0.0306$.  
While the sensitivity to numerical uncertainties is reduced here, more accurate evaluations of $v_\alpha$ will be required to produce a better estimate of the central charge $a$ with lower error using this approach. 

The calculations outlined above, among others, will help to test our ansatz regarding the structure of the corner coefficient $v_\alpha$.
If this ansatz is ultimately true, such entanglement calculations would produce a significant increase in the efficiency of determining the charge $a$, \eg
compared to calculations involving spherical geometries, which are notoriously difficult to study numerically. 
The coefficient $a$ is of particular interest here since this quantity decreases monotonically under renormalization group (RG) flows.\cite{CARDY1988749,Komargodski:2011vj,Casini:2017vbe}  
Numerical extraction of this universal charge from a local geometry would have important consequences in interacting theories, where analytical calculations in the continuum are unavailable.
Also in this direction, it may be possible to examine trihedral corners in holographic theories using numerical calculations analogous to those in Refs.~\onlinecite{Fonda:2014cca,Fonda:2015nma}. These and other
tests would further strengthen the body of evidence supporting our conjectures about the structure of the corner coefficient for general critical theories in 3+1 dimensions, or alternatively help to rule them out.

\begin{acknowledgments}   
We are thankful to A. Burkov and W. Witczak-Krempa for stimulating discussions. This research was supported by Perimeter Institute for Theoretical Physics. Research at Perimeter Institute is supported by the Government of Canada through the Department of Innovation, Science and Economic Development, and by the Province of Ontario through the Ministry of Research, Innovation and Science. Calculations were performed using the computing facilities of SHARCNET. RGM and RCM are supported in part by funding from the Natural Sciences and Engineering Research Council of Canada (NSERC). RGM is also supported by a Canada Research Chair. RCM is also funded by the Simons Foundation through the ``It from Qubit'' collaboration.  RGM and RCM thank the Galileo Galilei Institute for Theoretical Physics for 
hospitality and the INFN for partial support during the completion of this work. MG is supported in part by NNSFC Grants No.~11775022 and No.~11375026 and also by the China Scholarship Council.  MG also gratefully acknowledges the support of the Perimeter Institute Visiting Graduate Fellows program.
\end{acknowledgments} 


\bibliographystyle{apsrev4-1}
\bibliography{Bib_Entanglement}


\end{document}

%% file: Cube_angles.pdf_tex
\begingroup%
  \makeatletter%
  \providecommand\color[2][]{%
    \errmessage{(Inkscape) Color is used for the text in Inkscape, but the package 'color.sty' is not loaded}%
    \renewcommand\color[2][]{}%
  }%
  \providecommand\transparent[1]{%
    \errmessage{(Inkscape) Transparency is used (non-zero) for the text in Inkscape, but the package 'transparent.sty' is not loaded}%
    \renewcommand\transparent[1]{}%
  }%
  \providecommand\rotatebox[2]{#2}%
  \ifx\svgwidth\undefined%
    \setlength{\unitlength}{1059.11740124bp}%
    \ifx\svgscale\undefined%
      \relax%
    \else%
      \setlength{\unitlength}{\unitlength * \real{\svgscale}}%
    \fi%
  \else%
    \setlength{\unitlength}{\svgwidth}%
  \fi%
  \global\let\svgwidth\undefined%
  \global\let\svgscale\undefined%
  \makeatother%
  \begin{picture}(1,0.99271965)%
    \put(0,0){\includegraphics[width=\unitlength,page=1]{Cube_angles.pdf}}%
    \put(0.18266095,0.7804991){\color[rgb]{0,0,0}\makebox(0,0)[lb]{\smash{$\theta$}}}%
    \put(0.69945033,0.88341705){\color[rgb]{0,0,0}\makebox(0,0)[lb]{\smash{$\varphi$}}}%
    \put(0,0){\includegraphics[width=\unitlength,page=2]{Cube_angles.pdf}}%
    \put(0.8552561,0.11567643){\color[rgb]{0,0,0}\makebox(0,0)[lb]{\smash{$\ell_0$}}}%
    \put(0,0){\includegraphics[width=\unitlength,page=3]{Cube_angles.pdf}}%
    \put(0.2203708,0.15021898){\color[rgb]{0,0,0}\makebox(0,0)[lb]{\smash{$\ell$}}}%
  \end{picture}%
\endgroup%